\documentclass{WileyMSP-template}
\usepackage{textcomp, gensymb} 
\usepackage{fancyhdr} 
\usepackage{url}
\usepackage{xcolor}
\usepackage{soul,xcolor}
\usepackage{cite}
\usepackage{float}
\usepackage{comment}
\usepackage{amsmath}
\usepackage{upgreek}

\begin{document}

\title{Graphene Terahertz Devices for Sensing and Communication}

\maketitle


\author{Anna-Christina Samaha $^\dagger$},
\author{Jacques Doumani $^\dagger$},
\author{T. Elijah Kritzell},
\author{Hongjing Xu},\\
\author{Andrey Baydin},
\author{Pulickel M. Ajayan},
\author{Mario El Tahchi},
\author{Junichiro Kono*}

\vspace{12pt}
\vspace{12pt}

\begin{affiliations}
Anna-Christina Samaha and Prof. Mario El Tahchi\\
Laboratory of Biomaterials and Intelligent Materials, Department of Physics, Faculty of Sciences 2, Lebanese University, P.O. Box 90656, Jdeidet, Lebanon \\
annachristina.samaha@st.ul.edu.lb and mtahchi@ul.edu.lb\\
\vspace{12pt}
Jacques Doumani and T. Elijah Kritzell\\
Department of Electrical and Computer Engineering, Rice University, 6100 Main Street, Houston, TX 77005, USA\\
Applied Physics Graduate Program, Smalley--Curl Institute, Rice University, 6100 Main Street, \\Houston, TX 77005, USA\\
jd74@rice.edu and tek3@rice.edu\\
\vspace{12pt}
Hongjing Xu\\
Department of Physics and Astronomy, Rice University, 6100 Main Street, Houston, TX 77005, USA\\
hongjing.xu@rice.edu\\
\vspace{12pt}
Dr. Andrey Baydin\\
Department of Electrical and Computer Engineering, Rice University, 6100 Main Street, Houston, TX 77005, USA\\
Smalley--Curl Institute, Rice University, 6100 Main Street, Houston, TX 77005, TUSA\\
baydin@rice.edu\\
\vspace{12pt}
Prof. Pulickel M. Ajayan\\
Department of Materials Science and NanoEngineering, Rice University, 6100 Main Street, \\ Houston, TX 77005, USA\\
Smalley--Curl Institute, Rice University, 6100 Main Street, Houston, TX 77005, USA\\
pma2@rice.edu\\
\vspace{12pt}
Prof. Junichiro Kono\\
Department of Electrical and Computer Engineering, Rice University, 6100 Main Street, Houston, TX 77005, USA\\
Smalley--Curl Institute, Rice University, 6100 Main Street, Houston, TX 77005, USA\\
Department of Physics and Astronomy, Rice University, 6100 Main Street, Houston, TX 77005, USA\\
Department of Materials Science and NanoEngineering, Rice University, 6100 Main Street, \\ Houston, TX 77005, USA\\
Carbon Hub, Rice University, 6100 Main Street, Houston, TX 77005, USA\\
kono@rice.edu\\
\vspace{12pt}
$^\dagger$Equally contributing authors.

\end{affiliations}

\vspace{12pt}
\vspace{12pt}

\keywords{Graphene, 2D materials, Terahertz spectroscopy, THz wave modulation, Molecular sensing}

\justifying

\begin{abstract}

Graphene-based terahertz (THz) devices have emerged as promising platforms for a variety of applications, leveraging graphene’s unique optoelectronic properties. This review explores recent advancements in utilizing graphene in THz technology, focusing on two main aspects: THz molecular sensing and THz wave modulation. In molecular sensing, the environment-sensitive THz transmission and emission properties of graphene are utilized for enabling molecular adsorption detection and biomolecular sensing. This capability holds significant potential, from the detection of pesticides to DNA at high sensitivity and selectivity. In THz wave modulation, crucial for next-generation wireless communication systems, graphene demonstrates remarkable potential in absorption modulation when gated. Novel device structures, spectroscopic systems, and metasurface architectures have enabled enhanced absorption and wave modulation. Furthermore, techniques such as spatial phase modulation and polarization manipulation have been explored. From sensing to communication, graphene-based THz devices present a wide array of opportunities for future research and development. Finally, advancements in sensing techniques not only enhance biomolecular analysis but also contribute to optimizing graphene’s properties for communication by enabling efficient modulation of electromagnetic waves. Conversely, developments in communication strategies inform and enhance sensing capabilities, establishing a mutually beneficial relationship.

\end{abstract}

\section{Introduction}
\setstcolor{red}
Graphene is a distinctive two-dimensional (2D) material comprising a honeycomb lattice structure formed by a single layer of $sp^2$-bonded carbon atoms. This material has been the focus of intensive research due to its exceptional mechanical, electrical, optical, and thermal properties~\cite{novoselov2004electric,GeimEtAl2007NM,poot_nanomechanical_2008,BonaccorsoEtAl2010NP,li_graphene_2020,ghosh_extremely_2008,grapheneStatus}. Graphene is considered one of the most robust materials, with a tensile strength that is 200 times greater than that of steel~\cite{Mechanical_Changgu}. Additionally, it is an excellent conductor of electricity, with an electrical conductivity that is 1000 times greater than copper~\cite{CastroNetoEtAl2009RMP}. Furthermore, graphene has a thermal conductivity that exceeds diamond's~\cite{balandin_thermal_2011}. Due to these remarkable properties, graphene has the potential to revolutionize various fields, including electronics~\cite{schwierz_graphene_2010}, photonics~\cite{opticaldevicesreview}, energy storage~\cite{raccichini_role_2015}, and biomedicine~\cite{biomed_review}. 

\begin{figure}[hbt]
  \centering
  \includegraphics[scale=0.24]{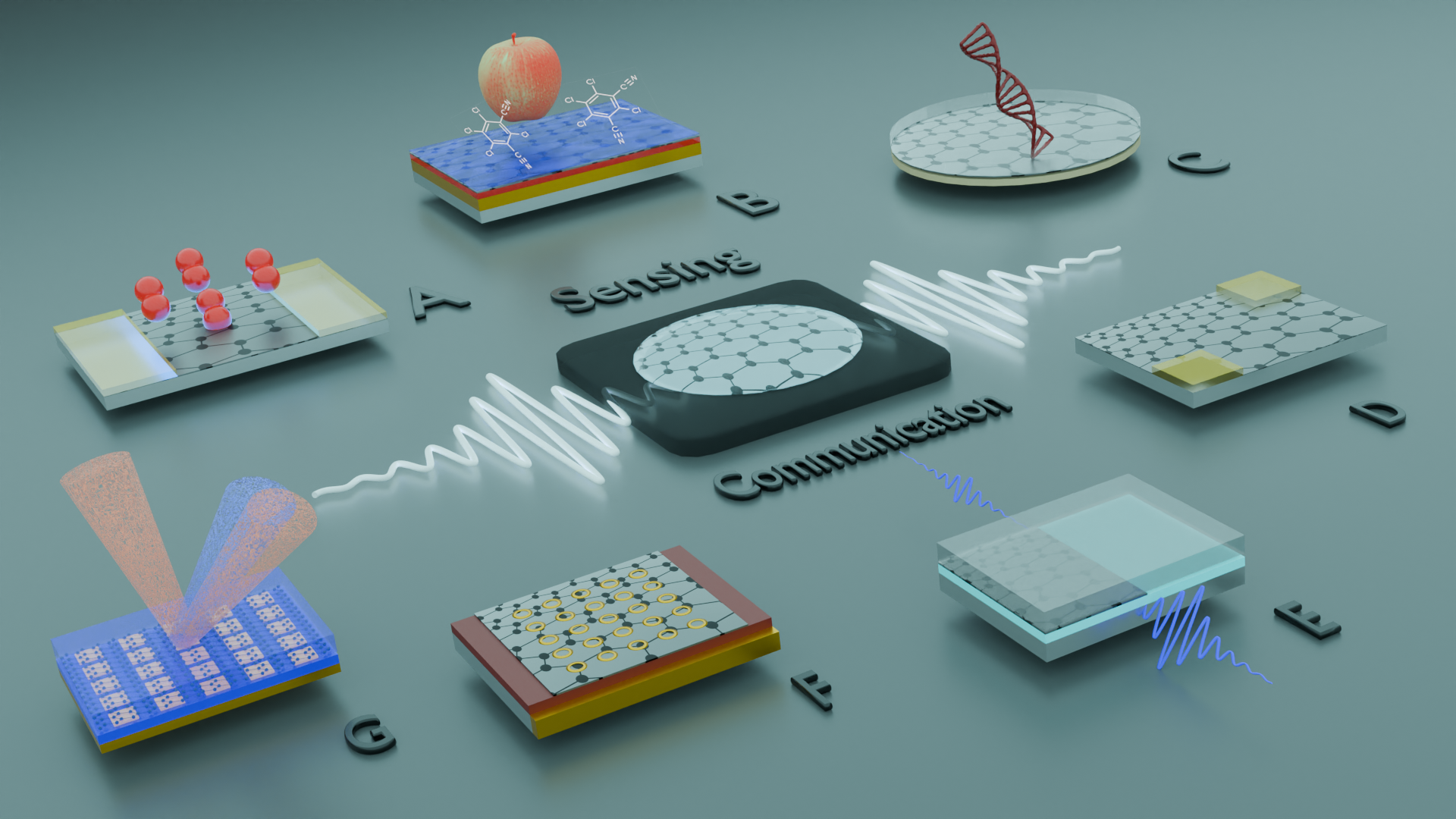}
  \caption{Various graphene-based THz devices discussed in this review for sensing and communication. The depicted devices are designed for: (A)~molecular sensing, (B)~pesticide sensing, (C)~DNA detection, (D)~absorption modulation, (E)~absorption enhancement, (F)~modulation enhancement, and (G)~phase modulation.}
  \label{fig:one}
\end{figure}


Due to the gapless Dirac-cone band structure, carriers in graphene exhibit ultrahigh electron mobilities and strong nonlinear response to electromagnetic (EM) fields~\cite{Hartmann_2014}.  Across an extensive spectral range, a myriad of optical phenomena have been explored in graphene to date: excitonic~\cite{excit}, plasmonic~\cite{grigorenko2012graphene,ni_fundamental_2018}, nonlinearity~\cite{nonlinearReview} including second-harmonic~\cite{shg}, third-harmonic~\cite{THG_new_tunable} and high-harmonic generation~\cite{HHG_with_elliptical_New}, Fano resonances~\cite{Fano}, and circular dichroism~\cite{kim_chiral_2016}.  In particular, linear gapless bands near the Dirac point allow excitation of charge carriers by low-energy photons, i.e., terahertz (THz) radiation~\cite{Hartmann_2014,nair2008fine,obraztsov2019coherent,zhou2023graphene}.

The THz frequency range, typically defined as the range between 0.1 THz and 10 THz, has attracted interest for sensing and imaging applications due to the ability of THz waves to penetrate various materials, thus providing spectral fingerprints for identifying molecular structures. For instance, macromolecules such as proteins and DNA exhibit their vibrational modes within this spectral band~\cite{nejad2016graphene}. Therefore, THz radiation is an invaluable tool for applications, ranging from chemical sensing to medical imaging~\cite{Mittleman:18}. Moreover, EM waves with THz frequencies hold significant promise for information and communication applications, including high-speed wireless communication~\cite{hasan2016graphene}. However, THz radiation falls between the microwave and infrared frequency bands, represented by the electronic and photonic regimes, respectively. The limited number of practical technologies for the generation and detection of THz radiation is commonly referred to as the ``THz technology gap.'' Developing efficient and tunable THz devices operating at room temperature remains challenging~\cite{hasan2016graphene}. Additionally, conventional materials often struggle to effectively respond to EM fields at THz frequencies, leading to significant propagation losses.

Graphene, with its distinct electronic properties mentioned above, is considered one of the most ideal candidates for filling the THz technology gap~\cite{Hartmann_2014}.
Both interband and intraband optical transitions contribute to the optical conductivity of graphene, but the latter is particularly significant at THz frequencies. Intraband transitions, or free carrier absorption, are essential for both the equilibrium and photoexcited states and significantly affect the material's response to THz radiation. Recent studies~\cite{Hartmann_2014,mak_measurement_2008,wang2016recent,mihnev_microscopic_2016,choi_broadband_2009,ren2012terahertz,tomadin_ultrafast_2018} have contributed crucial insights into the intricate dynamics of intraband transitions in graphene at THz frequencies, which lay the groundwork for exploiting graphene's capabilities in the development of devices for the generation, detection, and modulation of THz radiation~\cite{sensale-rodriguez_broadband_2012,li_graphene_2020,wang_dynamic_2022,zhou2023graphene, koppens_photodetectors_2014}. 

Hence, graphene is recognized as a pivotal material, offering unparalleled advantages for the advancement of THz technology, especially in sensing and communication. Compared to other materials, graphene stands out as an advantageous choice for fabricating sensors. This is attributed to its atomic thickness and high surface-to-volume ratio, which render it exceptionally sensitive to surrounding environmental changes~\cite{justino2017graphene}. Utilizing the intrinsic properties of graphene also presents a compelling solution for the development of tunable THz devices. As a 2D material, graphene can be easily integrated into various structures, making it a valuable asset in reducing optical losses~\cite{mittendorff2021optoelectronics}. Graphene also exhibits a high carrier mobility even at room temperature and can support surface plasmon polariton waves within the THz range~\cite{singh2020design}. The dynamics of these plasmons in graphene can help in minimizing losses during the transmission of THz waves~\cite{otsuji2012graphene}. Consequently, graphene emerges as a key component in THz communication systems, offering potential performance benefits over other THz devices.

Here, we explore the ways in which graphene is poised to fill the THz technology gap through pivotal advancements in sensing and THz wave modulation. First, we focus on molecular sensing applications. The electrical properties of graphene are sensitive to the adsorbed molecules and, therefore, result in changes to the THz optical conductivity. The adsorbed molecules can be detected via THz emission spectroscopy. Next, we review the states and dynamics of carriers in graphene as probed by THz time-domain spectroscopy, highlighting the Fermi-level-dependent response and low-density detection with THz parallel-plate waveguides. We also demonstrate some recent examples of graphene-based metamaterial systems that modulate the amplitude and phase of THz radiation.
Based on the knowledge accumulated in the literature, we conclude that graphene is poised to revolutionize THz technology, offering opportunities for innovation and advancement in both sensing and communication.

\section{THz Molecular Sensing with Graphene}

THz technology is emerging as a favorable approach for molecular sensing, providing advantages such as label-free and non-invasive detection~\cite{nsengiyumva2023sensing}. Molecules adsorbed on the surface of graphene can be identified based on their vibrational modes within the THz frequency region~\cite{wei2018application}. Therefore, graphene, characterized by its affordability, flexibility, and ease of transfer, holds significant potential as a material for fabricating devices dedicated to THz sensing applications~\cite{rangel2008graphene,lee2017nano,xu2019terahertz,C7NR03824K}. 

Of particular interest are advances in molecular adsorption detection and biomolecular sensing. Molecular detection presents notable opportunities across various domains. For instance, gas sensing is helpful in monitoring air quality, by identifying harmful emissions from sources such as transportation and industrial activities~\cite{buckley2020frontiers,demon2020graphene}. One example of a THz molecular sensor is the graphene heterogeneous silica D-shaped fiber, which leverages an electrically tunable frequency down-conversion process~\cite{an2020electrically}. Upon the adsorption of CO$_2$ gas molecules, for example, scattering impurities are introduced onto the graphene surface, and this changes graphene’s refractive index, enabling the detection of the gas~\cite{sun2016room}. THz molecular sensing is also favorable, compared to other frequencies, for H$_2$O-forming dynamics~\cite{Zhou2019Temperature}, investigation of flammable gases~\cite{Ngene2013Seeing}, and gas-metal interaction~\cite{Lee2023Advancements}.

Biosensing plays an important role in enhancing overall human well-being, particularly in fields such as biomedicine~\cite{shao2010graphene,heerema2016graphene,wekalao2023design}, gas detection~\cite{schedin2007detection,yuan2013graphene}, environmental monitoring~\cite{huang2023review}, and food safety~\cite{lang2023graphene}. Traditional methods, such as polymerase chain reaction and fluorescent microarrays, typically involve costly reagents and instruments, often lacking real-time monitoring capabilities~\cite{pena2018recent}. In contrast, biosensing offers rapid response times, making it an efficient alternative~\cite{yildiz2021graphene}. In the medical field, graphene-based THz biosensors offer the potential to detect diseases. For instance, a THz biosensor constructed with a graphene oxide framework enables interaction with various biomolecules~\cite{machado2022graphene}. In the case of an electrochemical biosensor, DNA can be adsorbed through non-covalent interactions~\cite{li2016biomolecules}, while proteins can adhere to graphene through both covalent and non-covalent interactions~\cite{li2016spontaneous}. A sensitive protein detection method for identifying breast cancer biomarkers can be achieved through the utilization of graphene-coated nanoparticles in a biomolecular sensor~\cite{myung2011graphene}.

\subsection{Molecular Adsorption Detection through THz Emission Spectroscopy}

\begin{figure} [hbt]
  \centering
  \includegraphics[scale=0.76]{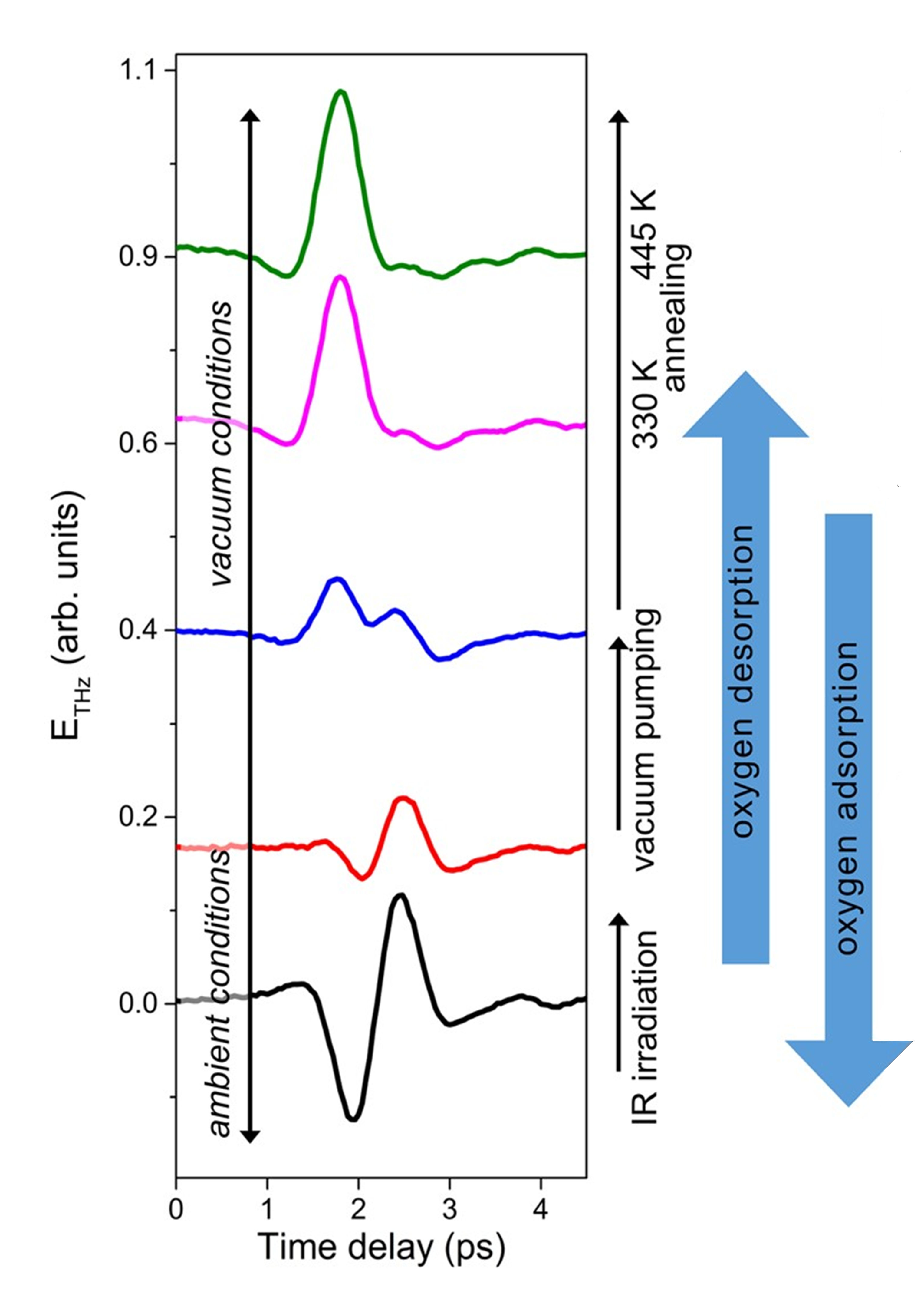}
  \caption{THz radiation waveform in the time domain emitted from graphene-coated InP under ambient conditions (in the presence of O$_2$ adsorbates), vacuum conditions and/or infrared irradiation, and annealing. Reproduced with permission under terms of Creative Commons CC BY license from \cite{bagiscan2017adsorption}. Copyright 2017, The Authors.}
  \label{fig:5}
\end{figure}


Atomically thin materials like graphene exhibit changes in electrical and optical properties due to interactions with their surroundings~\cite{santoso2014tunable, chang2014oxygen, docherty2012extreme, xing_regulating_2023}. For example, the adsorption of gas molecules, such as oxygen (O$_2$), which is recognized as a significant atmospheric gas adsorbate given its abundance, can affect material properties~\cite{bagiscan2017adsorption}. It has been shown that the transient THz photoconductivity and Fermi energy of doped graphene grown by chemical vapor deposition (CVD) are modulated in different gaseous environments, as demonstrated through optical pump--THz probe spectroscopy experiments~\cite{zhang_photoexcited_2017}. Understanding the environmental factors influencing graphene-based materials is thus important. THz time-domain spectroscopy (THz-TDS), especially THz emission spectroscopy~\cite{huang2019terahertz}, can be used to detect certain molecules and probe the dynamics of molecular adsorption and desorption. 

In the experiments conducted in \cite{bagiscan2017adsorption} and \cite{sano2014imaging}, a monolayer of graphene grown by CVD was placed on the surface of an indium phosphide (InP) crystal, known for its ability to emit coherent THz radiation efficiently. Specifically, when InP is photoexcited by femtosecond infrared pulses from a Ti:Sapphire laser, THz waves are generated from its surface. When the InP surface is coated by graphene, the impact of molecular adsorption and desorption on graphene can be probed through variations in the THz emission waveform under different conditions, acting as a molecular sensor.

\textbf{Figure \ref{fig:5}} shows that, under atmospheric conditions and ambient temperature, the generated THz waveform exhibits a dip at $\sim$2\,ps and a peak at $\sim$2.5\,ps. However, after the graphene-coated InP was subjected to vacuum pumping and/or continuous femtosecond infrared irradiation, which desorbs O$_2$, 
both peaks disappeared progressively, and a new positive peak replaced the dip at $\sim$2\,ps.
Further, when additional O$_2$ adsorbates were eliminated through annealing, along with any remaining water present at the junction of graphene and InP, the THz waveforms of graphene-coated InP and uncoated InP converged. Nevertheless, when the sample was left in the air for several hours without femtosecond laser excitation, O$_2$ molecules were re-adsorbed, leading to the recovery of the initial shape of the THz waveform. The original shape can also be restored by applying ultraviolet illumination to the sample, as it induces the oxidation of graphene~\cite{gunes2011uv, mitoma2013photo, zhao2012photochemical}. 
All these results imply that the variations in the emitted THz radiation waveform result from the adsorption and desorption of O$_2$ molecules on graphene.

\begin{figure}[hbt]
  \includegraphics[width=\linewidth]{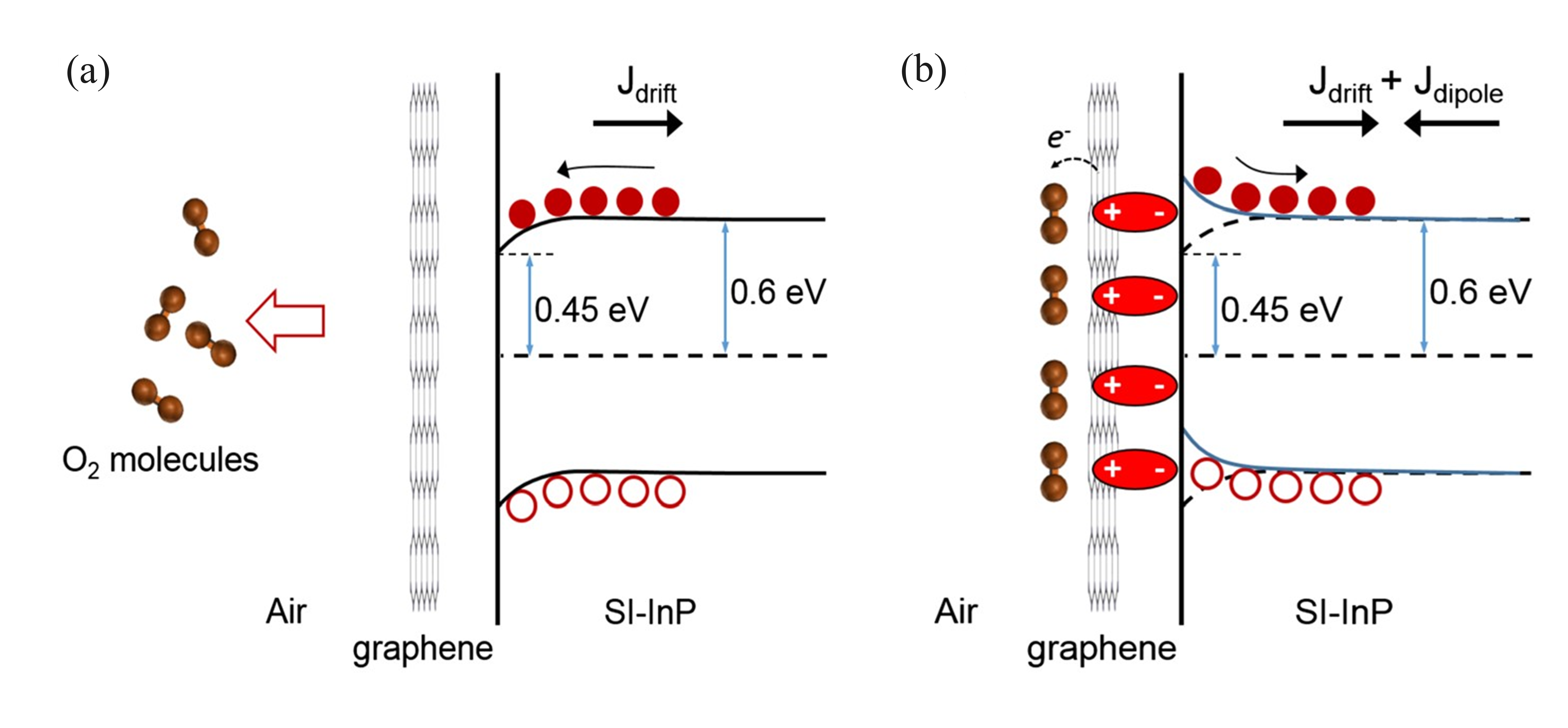}
  \caption{Band diagram of graphene/SI-InP explaining the principles of adsorption/desorption detection via THz emission spectroscopy. (a)~In the absence of O$_2$ adsorbates, the drift current $J_\mathrm{drift}$ flows toward the bulk of SI-InP. (b)~In the presence of O$_2$ adsorbates, the dipolar current $J_\mathrm{dipole}$ toward the surface is created. With a high concentration of O$_2$ adsorbates, the net surge current ($J_\mathrm{drift} + J_\mathrm{dipole}$) flows in the same direction as $J_\mathrm{dipole}$, toward the surface of SI-InP. Reproduced with permission from \cite{bagsican2016effect}. Copyright 2016, Springer Nature.}
  \label{fig:6}
\end{figure}

The effects of adsorption and desorption of O$_2$ molecules on the THz waveform discussed above can be understood by explaining the origin of the THz radiation from InP. The THz wave generation in semiconductors can happen through either nonlinear or linear processes~\cite{gu2002study}. In this study, the laser induced a low-intensity excitation, resulting in THz radiation primarily originated from linear processes that could be described by the current surge model at the semiconductor surface. The electric field magnitude, $E_\mathrm{THz}$, of the THz radiation is proportional to the time derivative of the transient current,
\begin{equation}
    E_\mathrm{THz} \propto \frac{\partial J(t)}{\partial t},
\end{equation}
as articulated in~\cite{bagiscan2017adsorption, sano2014imaging, nakajima2003polarity, nakajima2004competing}. Upon photoexcitating the sample, the current $J(t)$ is generated through two mechanisms: the acceleration of photoexcited carriers by the surface depletion field resulting from the laser excitation (drift current), and the photo-Dember effect caused by the difference in the diffusion velocities of electrons and holes (diffusion current)~\cite{huang2019terahertz,tonouchi2020simplified}. Given that InP is a semiconductor with a wide band gap ($E_\mathrm{g} = 1.34$\,eV), the contribution of the photo-Dember effect can be neglected. Therefore, the increase in current is primarily attributed to the transient photocarriers~\cite{gu2002study, bagsican2016effect}. The semi-insulating InP has a bulk Fermi level situated 0.6\,eV below the conduction band, whereas the surface Fermi level is fixed at 0.45\,eV below the conduction band~\cite{nakajima2003polarity, bagsican2016effect}. Consequently, the band curves down toward the surface, resulting in the drift current $J_\mathrm{drift}$ flowing in the opposite direction, toward the bulk of the InP substrate, as illustrated in \textbf{Figure \ref{fig:6}}(a) in the absence of O$_2$ molecules. On the contrary, when the sample is in contact with air, a transfer of electrons from graphene to O$_2$ adsorbates occurs~\cite{sano2014imaging, yang2011binding, levesque2011probing}, resulting in the formation of localized electric dipoles at the surface of InP, and the subsequent creation of an electric field. This electric field induces an upward bending of the InP band, leading to an additional current denoted as $J_\mathrm{dipole}$, which is in the opposite direction of the initial current $J_\mathrm{drift}$~\cite{du2022terahertz}; see Figure \ref{fig:6}(b). This alteration diminishes the amplitude of the emitted THz radiation from the sample. When there is a high concentration of O$_2$ adsorbates, $J_\mathrm{dipole}$ contributes more than $J_\mathrm{drift}$ to the net surge current ($J_\mathrm{drift} + J_\mathrm{dipole}$), which flows toward the surface in this case~\cite{sano2014imaging, bagsican2016effect}. This process explains the variation in the polarity of the THz waveform seen in Figure \ref{fig:5}.

While broad gas adsorbate sensing is critical across a number of fields, O$_2$ sensing is especially important for monitoring practical devices. Oxygen significantly alters semiconductor and metallic properties through doping~\cite{chang2014oxygen}. However, while well studied theoretically, effective characterization of O$_2$ adsorption has proved to be difficult~\cite{Zhou2015Mechanism, Liu2015Atomistic, Mehmood2013Adsorption,Lamoen1998Adsorption, Yan2012First, Guang2013First}. Typically O$_2$ detection is performed using transductive mechanisms~\cite{Ramamoorthy2003Oxygen} such as optical~\cite{Amao2009Optical}, photoacoustic~\cite{Cattaneo2006Photoacoustic}, or chemiresistance methods~\cite{Akbar2006High}, or mass uptake micro-electro-mechanical systems~\cite{Boisen2011Cantilever, Arshak2004Review, Dultsev2009QCM}. However, such devices are larger than monoatomic graphene, bear high fabrication costs, and rely on the chemical recognition of O$_2$ from other gas molecules. In comparison, graphene-enabled THz detection is passive and effective at characterizing the adsorbate. In Section~\ref{2.1}, with gating, we also report graphene's potential for studying N$_2$ gas adsorption and sensing. Graphene films and nanoribbons have also been shown to have high fidelity for detecting CO and NO$_2$ \cite{Joshi2010Graphene}.

\subsection{Biomolecular and Biochemical Sensing with Graphene}

\begin{figure} [hbt]
  \centering
  \includegraphics[scale=0.66]{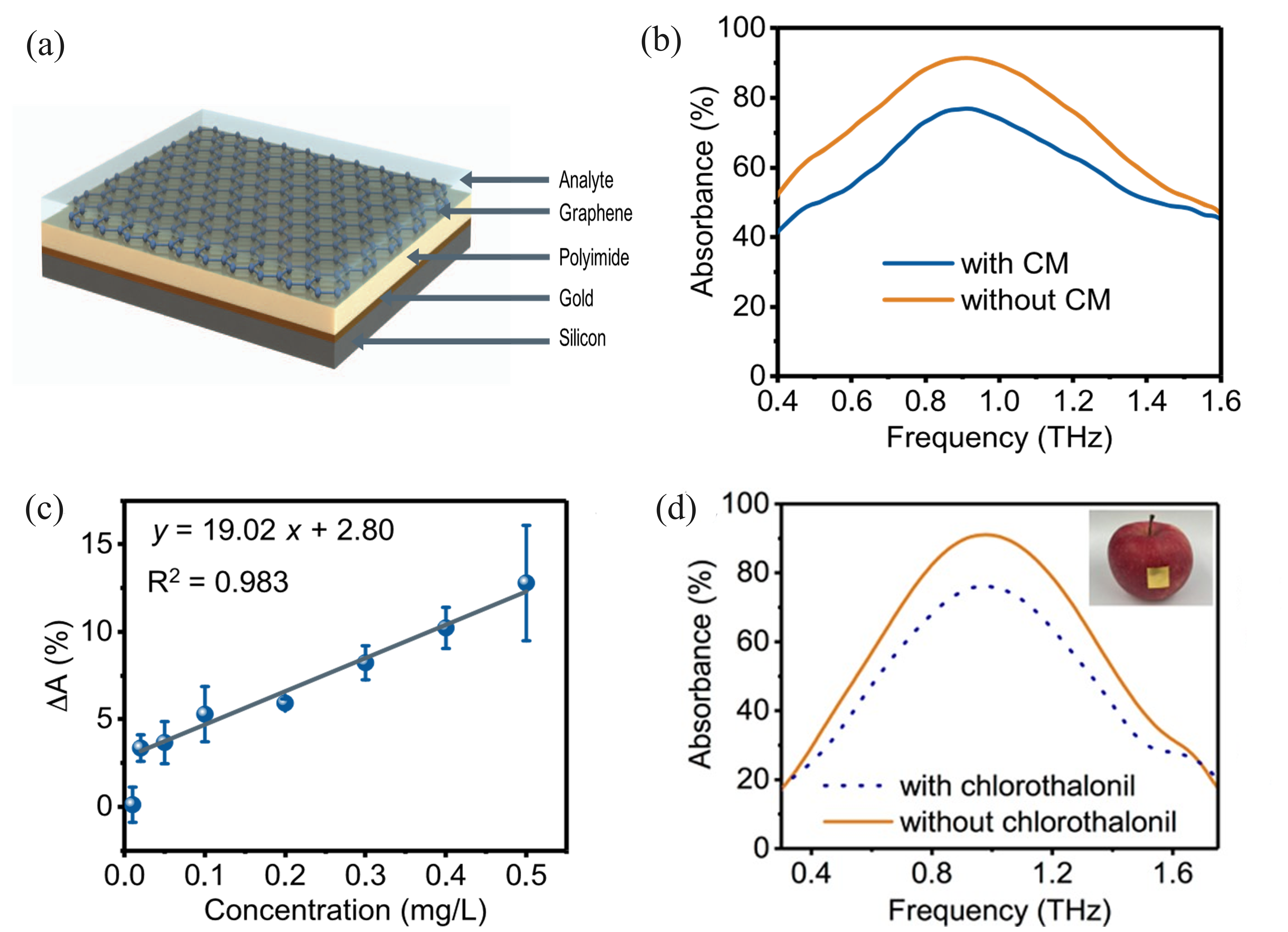}
  \caption{Biomolecular sensing with a graphene-based THz device. (a)~Schematic of the graphene sensor for detecting analytes. (b)~Absorption spectra for the graphene sensor with and without chlorpyrifos methyl. (c)~Absorbance variation with chlorpyrifos methyl molecules concentration ranging from 0.01 to 0.50\,mg/L. (d)~Absorption spectra for the graphene sensor with and without chlorothalonil. The inset displays a photograph of the flexible graphene sensor attached to the surface of an apple. Reproduced with permission from \cite{xu2020metamaterial}. Copyright 2020, American Chemical Society. }
  \label{fig:7}
\end{figure}

Information on the collective vibrational modes of biomolecules exists in the THz frequency range, and thus, THz spectroscopy is commonly used for chemical and biological sensing~\cite{ferguson2002materials}. The mismatch between the wavelengths of THz radiation and the dimensions of the target molecules limit the sensitivity of THz spectroscopy in detecting small amounts of analytes. To enhance the interaction between THz waves and analytes, and amplify the signal, metamaterials can be utilized. Metamaterials are easy to use in both reflection and transmission modes, making them valuable in the field of biosensors~\cite{sreekanth2016extreme, seo2020terahertz}. As the sensing device needs to be attached to a biointerface, flexibility is essential. However, the current processes for fabricating flexible metamaterials are expensive. Another issue is that the metamaterial structure introduces a gap separating the sensor and the target surfaces, diminishing the area of effective contact between the two and consequently reducing the sensing performance~\cite{xu2020metamaterial}. Hence, exploring methods to improve sensitivity without relying on metamaterials is underway, and cost-effective and sensitive metamaterial-free graphene-based THz molecular sensors have been proved promising. 

A promising approach to the enhancement of THz absorption in graphene involves a coherent-absorption-based trilayer structure. This structure consists of a layer of graphene and a metallic plate that are separated by a dielectric film, with the analyte of interest for investigation placed on top, as illustrated in \textbf{Figure \ref{fig:7}}(a). The dielectric material used is a polyimide film, and the metal layer is a gold back-reflector that redirects unabsorbed THz radiation back to the graphene sheet, preventing it from leaving the graphene sensor. This trilayer structure serves as a horizontal waveguide and confines the impinging light. Enhanced absorption occurs when the incident and the reflected waves are in phase as they reach the graphene layer~\cite{xu2020metamaterial}. Moreover, owing to the metallic nature of graphene, the structure satisfies the impedance matching requirement, ensuring alignment between the impedance of free space and that of graphene~\cite{watts2012metamaterial}. 

To examine the sensing capability of the graphene sensor, chlorpyrifos methyl (CM), a widely used pesticide~\cite{wei2012disposable}, can be used as an example of a biomolecule to detect. In Figure~\ref{fig:7}(b), THz absorption spectra are shown in the absence and presence of CM molecules at a concentration of 0.5\,mg/L. Initially, the absorption intensity is $\sim$91\%, but it decreases to $\sim$78\% with the introduction of CM molecules. This decrease is attributed to molecular adsorption, where the $\pi$ electrons of CM molecules encounter $\pi-\pi$ stacking interactions with graphene~\cite{xu2019terahertz,xu2022defect}. These interactions cause the Fermi level to move toward the Dirac point of graphene, leading to a reduction in carrier density~\cite{xu2020metamaterial, guo2013graphene}. The altered Fermi level increases the impedance mismatch between free space and the THz graphene sensor, resulting in a decrease in the absorption peak~\cite{tasolamprou2019experimental}. In Figure~\ref{fig:7}(c), the absorbance is presented as a function of CM concentration (0.02 to 0.50\,mg/L), which shows good linearity with a coefficient of determination $R^2 = 0.983$. The limit of detection is determined to be 0.13\,mg/L~\cite{xu2020metamaterial}, which is better than that obtained using metamaterials standalone (0.204\,mg/L)~\cite{xu2017terahertz}. This technique could be extended for the sensing of other pesticides such as chlorpyrifos methyl molecules on rice~\cite{D2RA06006J} and chloramphenicol in milk~\cite{su_terahertz_2024}.

\begin{figure} [hbt]
  \centering
  \includegraphics[scale=0.785]{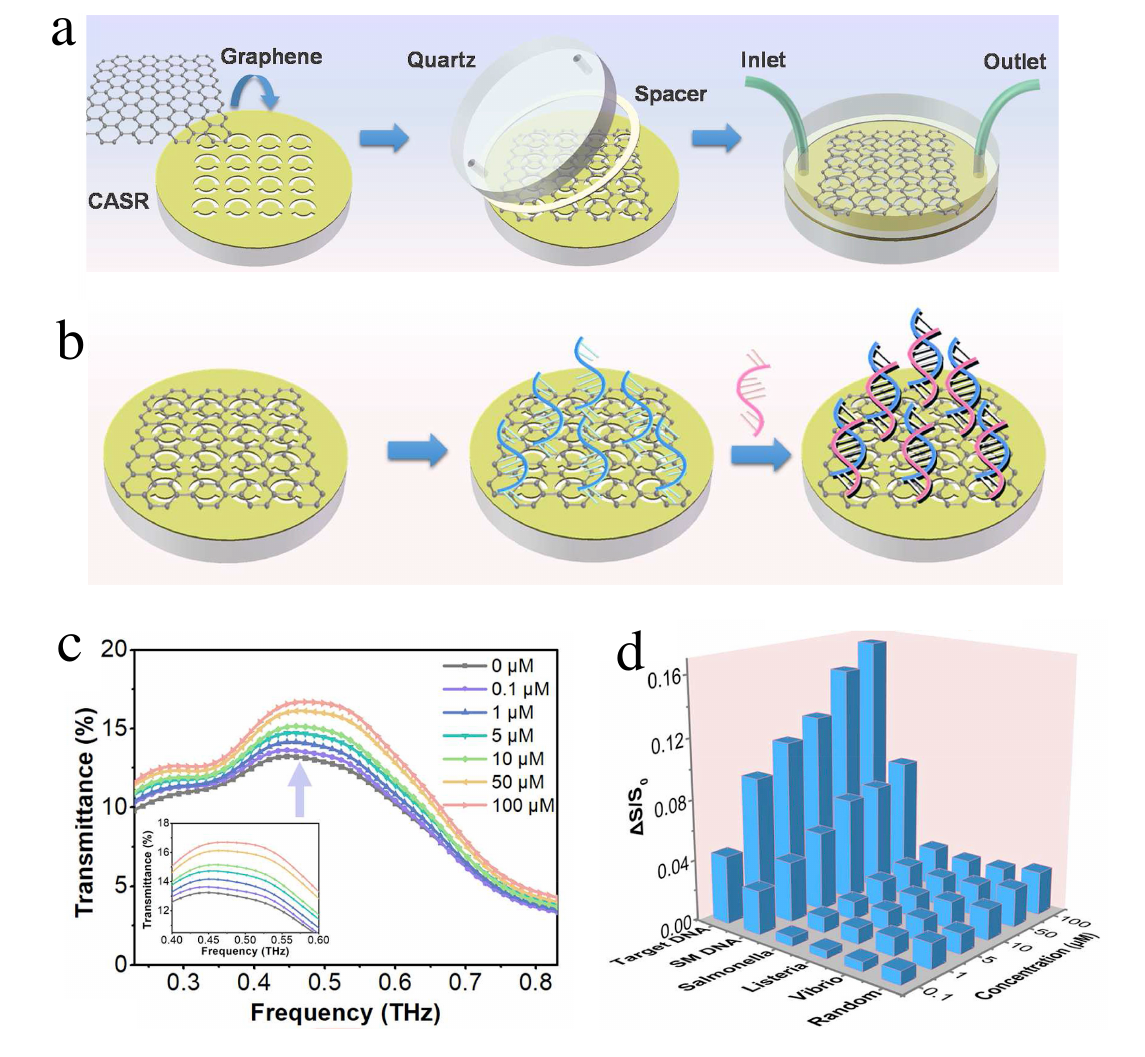}
  \caption{DNA-based THz-sensing with CASR-graphene TMFC assembly. (a)~Schematics depicting device fabrication. (b)~Schematics illustrating the core part of the TMFC. DNA probe (blue strands) is introduced to immobilize/hybridize the target (red strands). (c)~THz transmission spectra for target DNA (Eae gene sequences of Escherichia coli O157:H7) at various injected molar concentrations. (d)~3D specificity and sensitivity histogram of various target DNA species. Reproduced with permission from \cite{DNA}. Copyright 2021, Elsevier.}
  \label{DNAfig}
\end{figure}

Furthermore, the fabrication of a flexible graphene sensor suitable for utilization at biointerfaces for pesticide sensing can be readily achieved by using electrically conductive tape as the metal layer in the trilayer structure. The resulting flexible graphene sensor can adhere to a curved surface without any gaps. For instance, when detecting chlorothalonil molecules on an apple surface at a concentration of 0.60\,mg/L [Figure~\ref{fig:7}(d)], a reduction in the absorption peak is detected in the presence of chlorothalonil, similar to the previous case. This demonstrates the efficacy of this flexible graphene-based THz sensor in detecting pesticides at a biointerface~\cite{xu2020metamaterial}. 

Innovative approaches to enhance adsorption detection sensitivity involve metamaterial structures~\cite{xie_extraordinary_2015}, such as nanoslots with tuned dimensions for resonance. Patterning of graphene either before or after metamaterial assembling has well been explored~\cite{s23135902}. Mediated graphene nanoslot assemblies have demonstrated amplified sensitivities for detecting various single-stranded DNA species at the nanomole level, while patterned graphene nanoantennas enhanced sensitivity for lactose detection~\cite{DNA}. Another study highlights virus detection capabilities through polarization-state sensing using a graphene chiral metasurface~\cite{virus}.


A new technological advancement has been enabled, allowing for an \text{in~situ}, non-invasive, high-sensitivity, high-specificity DNA sensing of liquid samples. Most biological samples exist in liquid form, and water is known for strongly absorbing THz radiation~\cite{AHMADIVAND2020108,ALFIHED2020112393}, limiting THz-sensing capabilities to only solid/dried samples. In a new twist, a THz microfluidic cell (TMFC) has been developed for enabling THz-sensing on liquid samples at low volumes, bypassing any distorting absorption [\textbf{Figure~\ref{DNAfig}}(a)]. First, CVD-grown graphene is transferred onto a complementary asymmetric split-ring (CASR) Au-metasurface for enhancing THz absorption. The rest of the TMFC assembly is joined by adding a quartz window with two holes, built with a spacer to realize the microchannel. Then probe DNA is introduced into TMFC as an aptamer that immobilizes/hybridizes target DNA on CASR [Figure~\ref{DNAfig}(b)].

THz-TDS spectroscopy reveals that the transmittance increases with target DNA, Eae gene sequences of Escherichia coli O157:H7, and concentration increases. The concentration was varied between $0.1$ and 100\,$\upmu\mathrm M$, as presented in Figure~\ref{DNAfig}(c). Furthermore, the sensing performance of such devices is characterized by the relative change rate $\Delta S \text{/} S_\mathrm{0} = (S_\mathrm{DNA}-S_\mathrm{Water}) \text{/} S_\mathrm{Water}$, with the effective transmission area $S = \int_{f1}^{f2} T(f) \, df$, quantifying the transmission strength in a frequency range bin, $f$ being the frequency, and $T$ the transmittance. In Figure~\ref{DNAfig}(d), CASR-graphene TMFC biosensors demonstrate not only high sensitivity DNA detection at low concentrations of 100\,nM, but also high specificity to target DNA for testing, which is observed with higher effective transmittance area's ratio change compared to the other introduced dummy DNA species. The authors discuss the possibility of further improving the sensing capabilities by incorporating deep learning to extract minute transmission variations, two layers of graphene on CASR for enhanced signal acquisition, metasurfaces for probing highly confined EM modes. Finally, they discuss the advantages of incorporating such technology with other materials such as carbon nanotube films aligned with the vacuum filtration technique~\cite{HeetAl16NN,doumani_engineering_2023,doumani_macroscopically_2023} and smart hydrogels~\cite{ZHOU2021122213}, for simplifying the fabrication, lowering the fabrication and production costs, and better scalability and device stability.


\section{THz Wave Modulation with Graphene}

Understanding carrier dynamics in graphene is paramount for its optoelectronic applications. Ultrafast pump-probe spectroscopies, time-resolved Raman spectroscopy, and measurements of photocurrent have all been useful in revealing insights into electron-electron, electron-phonon, and phonon-phonon interactions~\cite{park2009imaging,kang2010lifetimes,breusing2011ultrafast,sun2008ultrafast,newson2009ultrafast,ruzicka2012spatially,lin2013ultrafast,wang2010ultrafast,golla2017ultrafast,malard2013observation,gatamov2020fluence, scarfe2021systematic}. Both intraband and interband contributions to the optical conductivity of graphene have been probed via a variety of steady-state and transient spectroscopy methods~\cite{dawlaty2008measurement,george2008ultrafast,shang2010femtosecond,ruzicka2010femtosecond,huang2010ultrafast,shang2011ultrafast1,huang2011ultrafast,zhao2011ultrafast,breusing2011ultrafast,tani2012ultrafast,chen2014diversity,kadi2014microscopic,chen2016non,tomadin_ultrafast_2018,gatamov2020fluence,guo2020control, pistore2022mapping,nguyen2023non}. 
THz spectroscopy, in particular, plays a crucial role in characterizing carrier dynamics, offering deep insights into the properties of materials~\cite{lu2018critical,ivanov2015perspective,docherty2012terahertz,mihnev_microscopic_2016, jnawali2013observation, frenzel2013observation,singh_terahertz_2022}. This sensitive approach not only enhances our understanding of carrier behavior but also contributes to the development of next-generation wireless communication devices.

While research into the necessary components for constructing THz communication networks with high bandwidth remains limited, 2D materials have been proposed as potential candidates for THz wireless communication devices. With its high conductivity, graphene is considered as a favorable option for communication applications ~\cite{sensale2011unique,abohmra2022two,taghvaee2022multi}.
The following paragraphs discuss the modulation of the Fermi energy using gate voltage control. Next, the enhancement of THz absorption is examined, employing parallel-plate waveguides and total internal reflection geometries to achieve large-bandwidth tunable absorption. Next, we explore the enhancement of THz gated-modulation, notably through the use of ring-shaped apertures, which exhibit the extraordinary optical transmission effect. Finally, the discussion is shifted to phase modulation via gating and employing metasurfaces.

\subsection{Gate Control of THz Wave Transmission}\label{2.1}

\begin{figure}[hbt]
  \centering
  \includegraphics[scale=0.74]{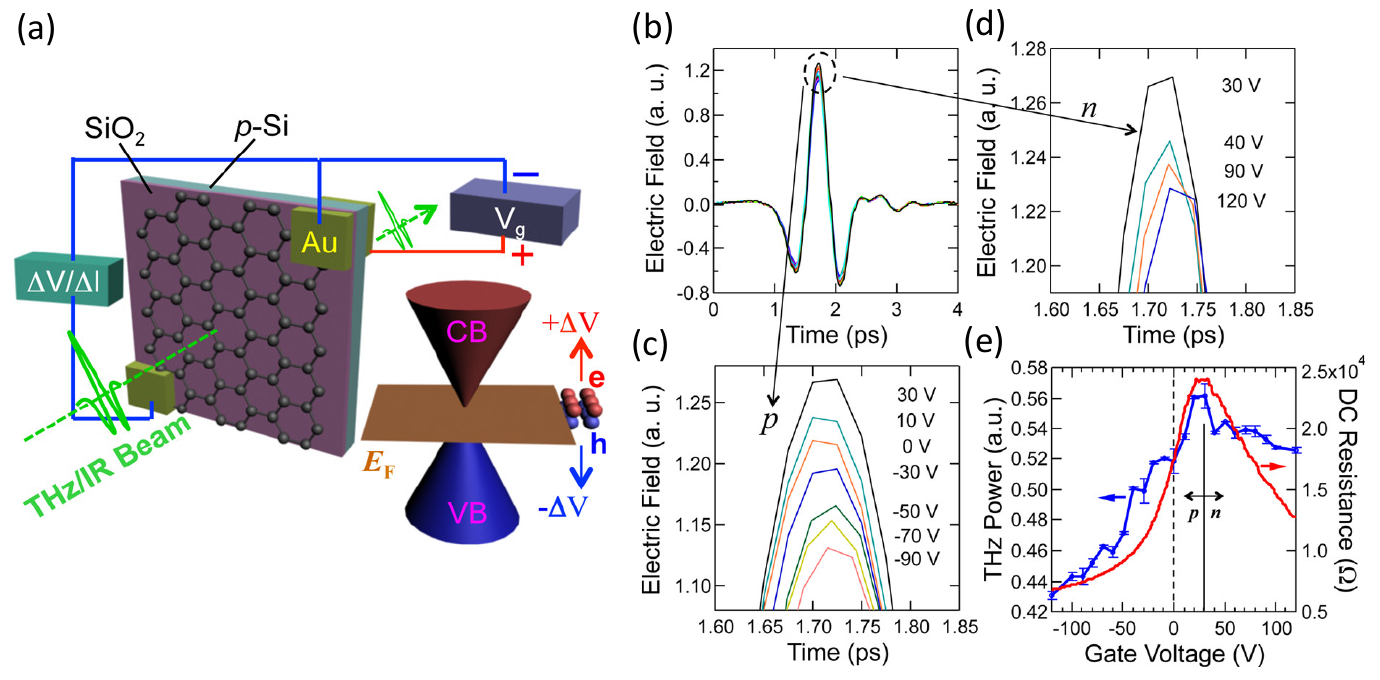}
  \caption{THz time-domain spectroscopy on graphene with gate-voltage-tuned Fermi energy, $E_\mathrm{F}$. (a)~Illustrative depiction of the graphene device with gating over a substantial area and the band dispersions of graphene on the right. (b)~Representation of THz waveforms in the time domain after transmission with varying gate voltages from $-$90 to 120\,V, focused on (c)~$p$-doped and (d)~$n$-doped regimes at the maximum THz electric field. (e)~The blue line with \st{open} circles represents THz transmission power, and the red solid line represents DC resistance, plotted against the gate voltage, showing a neutrality point at 30\,V. Reproduced with permission from \cite{ren2012terahertz}. Copyright 2012, Springer Nature.}
  \label{fig:1}
\end{figure}

Since the position of the Fermi energy, $E_\mathrm{F}$, influences intraband and interband transitions~\cite{tomadin_ultrafast_2018, mihnev_microscopic_2016}, carrier-density-dependent spectroscopy is particularly useful. The changes in carrier density can be sensitively monitored through THz-TDS or Fourier-transform infrared (FTIR) spectroscopy~\cite{cinquanta2023charge,han2020time, horng2011drude}. 
One way to tune $E_\mathrm{F}$ is through chemical doping, but replacing carbon atoms in graphene with other elements suppresses the mobility~\cite{lee2018review,gosling2021universal}. Consequently, the most effective method for tuning $E_\mathrm{F}$ is by implementing a variable gate voltage, $V_\mathrm{g}$.

\textbf{Figure~\ref{fig:1}}(a) shows a large-area (centimeter-sized) field-effect transistor, consisting of monolayer graphene on a SiO$_2$/$p$-Si substrate~\cite{ren2012terahertz, gao2014terahertz}. The electrodes laying on top of graphene and on the backside of the $p$-Si substrate, respectively, are connected to a DC voltage supply, and simultaneously, a lock-in amplifier is employed to monitor the DC conductance (or resistance) of the graphene layer~\cite{ren2012infrared}. A THz EM wave is normally incident onto the sample, and its transmitted intensity changes based on the gate voltage, $V_\mathrm{g}$, applied between the graphene layer and the Si substrate. As depicted in Figure~\ref{fig:1}(b), the highest THz transmission is obtained at $+$30\,V when the Fermi energy reaches the charge neutrality point. For voltages below and above this value, there is a monotonic decrease in THz transmission with the voltage variation, as illustrated in Figure~\ref{fig:1}(c-d). Figure~\ref{fig:1}(e) shows the transmitted THz beam power (blue circled line) alongside the DC resistance (red trace) versus $V_\mathrm{g}$. A maximum DC resistance is observed at $+$30\,V, demonstrating conformity with the gate dependence of the transmitted THz power, which peaks at this voltage. Thus, this work clearly demonstrates that gated large-area graphene samples are useful for manipulating THz transmission by tuning $E_\mathrm{F}$~\cite{ren2012terahertz, ren2012infrared}.
Similarly, Wu and coauthors have reported a graphene/ionic liquid/graphene sandwich structure that modulates a broadband frequency range from 0.1 to 2.5\,THz with a small gate voltage~\cite{wu_graphene_2015}.

\subsection{Enhancing THz Absorption}\label{2.3}

\begin{figure}[hbtp]
  \centering
  \includegraphics[scale=0.63]{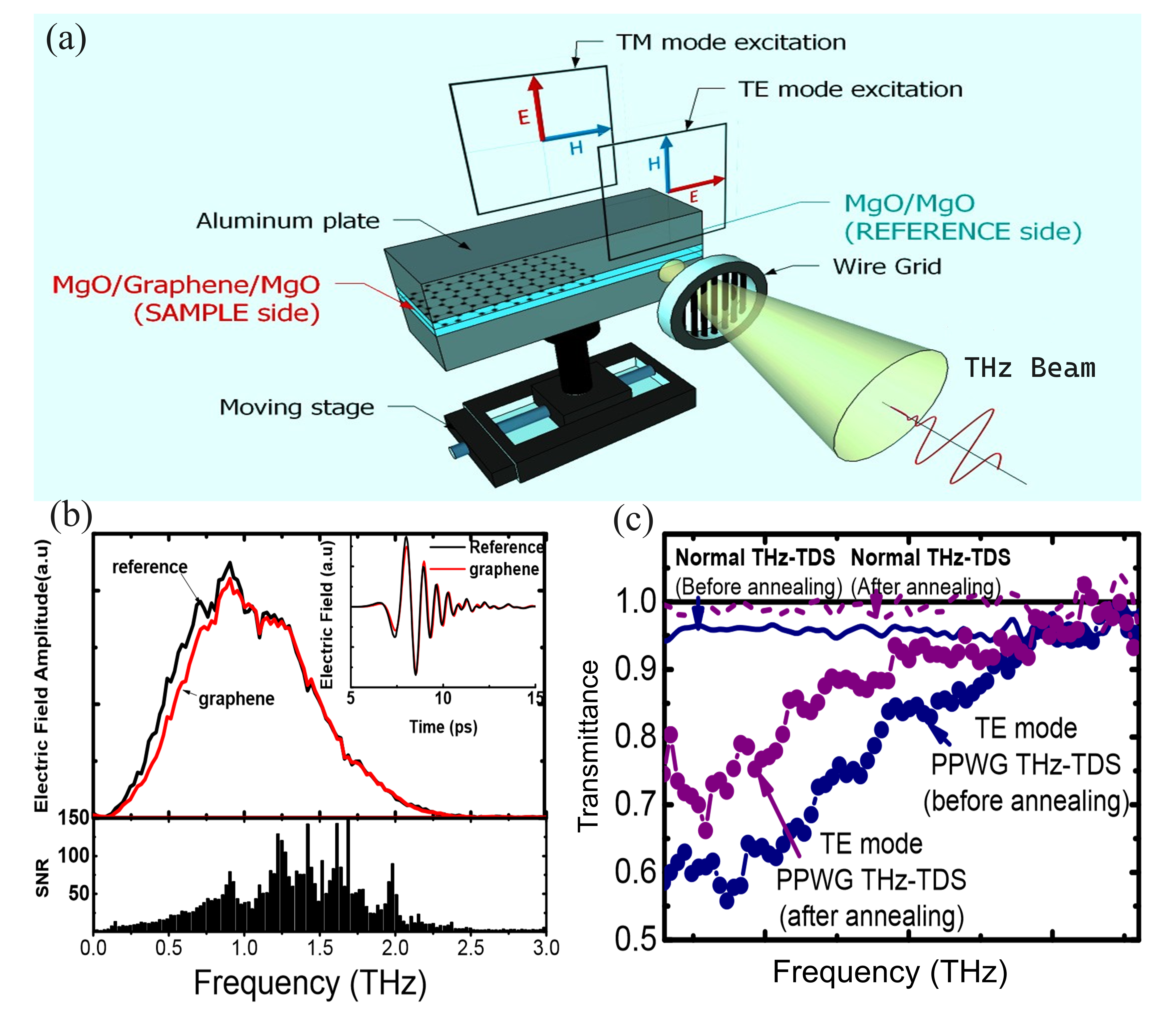} 
  \caption{PPWG setup and THz-TDS transmission spectra in the TE and TM modes. (a)~PPWG configuration with graphene sandwiched between two THz transparent MgO substrates. The assembly is sandwiched between two aluminum plates. Graphene is stretched across one portion of the device making space for a reference. The THz beam polarization is controlled by the wire grid. The PPWG system is placed on a translation stage that moves it in small steps to the left and right of the incident wave propagation direction. (b)~THz spectra for the graphene sample (red curve) and the reference (black curve) are in the upper panel, with the associated SNR spectrum in the lower panel. The inset displays the corresponding TE mode time-domain THz waveforms. The small dips result from the residual water vapor. (c)~Transmission spectra of the conventional THz-TDS and the PPWG THz-TDS for the graphene sample in TE mode, before and after annealing. For the conventional THz-TDS, the non-annealed graphene is represented by the blue solid curve, and the annealed graphene by the purple dashed curve. The same sample is characterized using the PPWG THz-TDS; the blue solid circles indicate transmittance before annealing while the purple solid circles indicate transmittance after the annealing process. Reproduced with permission from \cite{razanoelina2016probing}. Copyright 2016, Optical Society of America.}
  \label{fig:3}
\end{figure}

\begin{figure} [hbtp]
  \centering
  \includegraphics[scale=0.76]{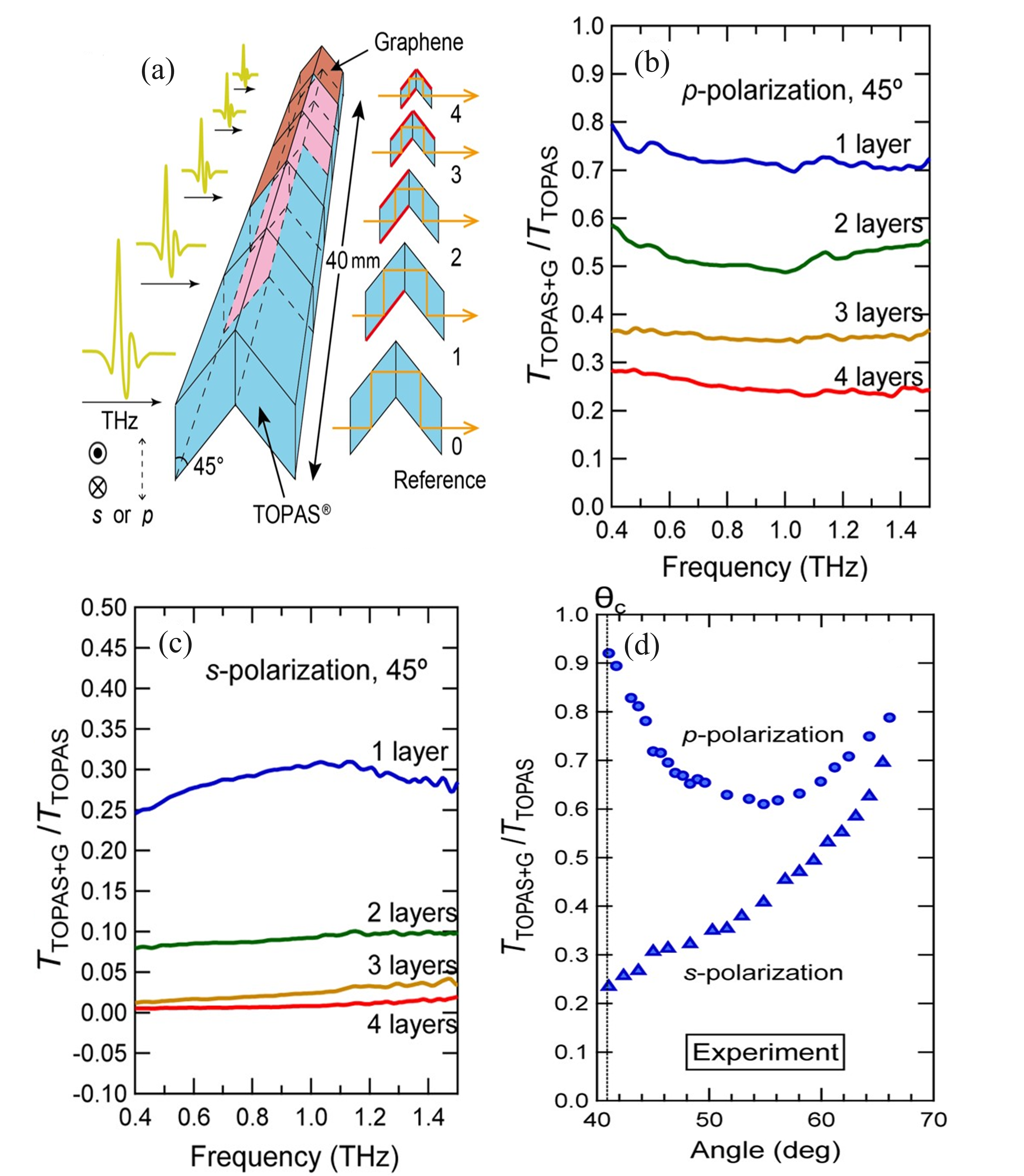}
  \caption{Graphene THz wave absorption enhancement through total internal reflection. (a)~Representation of the waveguide geometry of graphene-on-TOPAS with different numbers of graphene reflections. (b-c)~Relative transmittance spectra for a $p$- ($s$-) polarized THz wave at a 45$^\circ$ angle of incidence, respectively. (d)~Variation in THz transmittance with angle for both polarizations. The angle of incidence on graphene is adjusted using a rotatable stage placed under the waveguide. Reproduced with permission from \cite{harada2017giant}. Copyright 2017, American Chemical Society.}
  \label{fig:4}
\end{figure}

Studying the THz response of thin materials like monolayer graphene can be challenging due to their short interaction length with THz radiation. As discussed in Section~\ref{2.1}, increasing the carrier density by applying a gate voltage to graphene can enhance its response to THz radiation~\cite{ren2012infrared}, but this technique necessitates gate electrodes and is not always applicable to common 2D materials. Moreover, the influence of gate-induced carriers on the substrate must be considered. 

Another way to achieve a better response is to insert graphene into a parallel-plate waveguide (PPWG). This configuration provides a longer interaction length than the conventional THz spectroscopy, resulting in higher sensitivity. Additionally, this method facilitates the detection of the behavior of 2D materials in response to diverse polarizations of the EM field through waveguide mode control~\cite{razanoelina2015parallel, razanoelina2016probing, wu2021fano}.

A waveguide structure typically used in the PPWG scheme~\cite{razanoelina2016probing} is illustrated in \textbf{Figure \ref{fig:3}}(a), which consists of two aluminum plates. In the first half of the PPWG, CVD-grown graphene is inserted between two THz-transparent MgO substrates. In the other half, bare MgO substrates are used as a reference. A linear polarizer is set before the waveguide to provide $s$- or $p$-polarization configurations, allowing the excitation of either the transverse electric (TE) or transverse magnetic (TM) mode~\cite{razanoelina2015parallel,razanoelina2016probing}. 

THz spectra obtained for both the reference and the graphene sample are illustrated in Figure~\ref{fig:3}(b), with the THz wave polarized in the TE mode, while the inset shows the corresponding time-domain THz waveforms. For PPWG THZ-TDS, the alignment of the waveguide greatly contributes to the accuracy of the data. 
Therefore, THz pulses for the sample and the reference are recorded for ten different locations each. 
The signal-to-noise ratio (SNR) is calculated by dividing the average amplitude of the THz spectrum by its standard deviation~\cite{naftaly2009methodologies}. 
Figure~\ref{fig:3}(c) presents a comparison of the transmittance spectra between conventional THz-TDS and PPWG THz-TDS for monolayer graphene in TE mode, before and after annealing. In conventional THz-TDS, the spectra remain almost flat across the frequency range, with transmittance values of $\sim$95\%, both before and after annealing. On the other hand, the spectra obtained with PPWG THz-TDS show variations in transmittance depending on the frequency. In the low-frequency regions, the transmittance is $\sim$60\% before annealing, and $\sim$70\% after annealing. These values increase to $\sim$95\% at the end of the spectra for high frequencies. Hence, using PPWG THz-TDS results in a $\sim$30\% decrease in transmission, successfully achieving THz absorption enhancement in graphene despite its low-density carrier ($\sim$2~$\times$~10$^{11}$~cm$^{-2}$). This underscores the advantage of this approach over the conventional method, which was blind to any substantial carrier density change. This result is significant in future wireless communication systems, considering graphene's broadband optical absorption~\cite{hasan2016graphene,liu2011graphene}.
In TM mode, the electric field is perpendicular to the graphene layer, implying no induced current within the graphene. Experimental results indicate that the electric field amplitude measured for the sample matches that of the substrate across the spectrum, suggesting no interaction or absorption between the graphene layer and incident waves.

An alternative method for enhancing THz absorption in graphene is using a total internal reflection (TIR) geometry~\cite{sun2020exploiting}, where the absorption of electromagnetic EM waves can be tuned between 100\% and 0\% by changing the $E_\mathrm{F}$ of graphene~\cite{ukhtary2015fermi,harada2017giant}. Enhanced modulation depth in the THz regime has been achieved using a wire grating and ionic gel-based graphene TIR THz modulators~\cite{liu2017graphene, sun2018graphene}. The geometry used in \cite{harada2017giant} consists of a large-area graphene sample inserted between two dielectric media.  One of the chosen dielectric media is thermoplastic olefin polymer of amorphous structure (TOPAS), with a refractive index ($n_1$ = 1.523 $\pm$ 0.002) and a negligible absorption across the THz range~\cite{dangelo2014ultra,rahmanshahi2021tunable,ye2019composite}, while the other is air ($n_2$ = 1). To satisfy the TIR condition, the THz wave must have an angle of incidence $\theta$ greater than the critical angle $\theta_\mathrm{c}$ ($\theta > \theta_\mathrm{c}$), and according to Snell's law, $\theta_\mathrm{c} = \sin^{-1}(n_2/n_1)=41.06^\circ$.

The design of the ``graphene-on-TOPAS'' waveguide is schematically shown in \textbf{Figure \ref{fig:4}}(a). Two parallelogram-shaped TOPAS are combined to form a prism with six TOPAS/air interfaces. While two of the facets are left graphene-free (one for receiving the incident THz beam and the other for transmitting it), a specific number of the remaining four interfaces are covered with graphene. This is done to measure the transmittance for cases involving $n =$ 0, 1, 2, 3, and 4 reflections by graphene, using a linear translation stage to move the waveguide perpendicularly to the direction in which the beam is propagating. The parallelogram faces of the prism have one angle equal to 45$^\circ$ to get an incident angle of $\theta =$ 45$^\circ$ when the THz beam is normally incident on the uncovered interface. The transmission signal without reflection by graphene is used as reference to calculate the relative transmittance $T_\mathrm{TOPAS+G}/T_\mathrm{TOPAS}$, where $T_\mathrm{TOPAS+G}$ represents the transmittance of the waveguide with graphene ($n =$ 1, 2, 3, or 4) and $T_\mathrm{TOPAS}$ without graphene ($n =$ 0).

Transmission measurements are carried out for both $p$- and $s$-polarized THz beams. Given that the THz wave is linearly polarized, the waveguide structure can be rotated by 90$^\circ$ in a plane orthogonal to the incident beam to obtain either $p$- or $s$-polarization. Relative transmittance spectra for varying numbers of graphene layers are shown in Figure~\ref{fig:4}(b) for $p$-polarized THz waves and in Figure~\ref{fig:4}(c) for $s$-polarized THz waves. After performing spectral averaging within the 0.5-1.5~THz range, in the case of $p$-polarization, graphene absorbs 28.2\% of the incident wave with every reflection, leaving $(1-0.282)^4 = 26.2\%$ after four reflections. In contrast, for $s$-polarized THz waves, graphene absorbs 71.0\% of the incident wave with every reflection, resulting in just $(1-0.710)^4 = 0.7\%$ after four reflections. These results indicate high THz absorption, particularly for $s$-polarization.


On the other hand, the variation in graphene absorption with the incident angle is studied by rotating one-parallelogram TOPAS geometry in a parallel plane to the beam, for the purpose of changing the angle $\theta$. The spectrally averaged transmittance $T_\mathrm{TOPAS+G}/T_\mathrm{TOPAS}$ is measured as a function of $\theta$ for both polarizations. As seen in Figure~\ref{fig:4}(d), starting with the critical angle $\theta_\mathrm{c}$, the transmittance in the case of $p$-polarization is high, indicating minimal absorption by graphene. As the angle increases, the transmittance decreases until it reaches 0.61 at an angle $\theta=55^\circ$. After this angle, the transmittance increases again to approximately 0.8 at $66^\circ$. The transmittance is 23.5\% for $s$-polarization at the critical angle $\theta_\mathrm{c}$, and this value increases with the angle $\theta$ until it reaches $\sim$70\% at $66^\circ$. Hence, both polarizations depend on the incident angle, and this dependence differs significantly between them.

\subsection{Enhancing THz Wave Modulation}
Integrating gated graphene into THz devices enables manipulation of its electronic characteristics and achieves high-speed optical modulations for THz communications and imaging~\cite{hasan2016graphene,kim2023highly,zaman2022terahertz,ma2019modulators, Li2022gate, Kim2017Amplitude, Miao2015Widely, Chen2018Graphene, Delinnocenti2014Low}.
Graphene-based THz modulators tend to display limited on/off ratios due to the single-layer thickness of graphene and the absence of resonance in intraband absorption within the THz region~\cite{gao2014high}. Enhancement of the extinction ratio and THz absorption of the graphene-engineered modulator can be achieved by using subwavelength apertures in a conductive medium that exhibit the extraordinary optical transmission (EOT) effect~\cite{chen2008electronic, seo2009terahertz, hendry2007ultrafast, lee_switching_2012, cao2021multi,gao2021tunable}. Ring-shaped apertures are particularly effective due to their strong polarization-insensitive EOT effect~\cite{shu2011high,fu2022polarization,gao2021artificial}. EOT describes a phenomenon in which a structure featuring subwavelength apertures in a metallic film enables greater light transmission than what is anticipated by ray optics. This occurs due to an enhancement of the electric field inside and near the apertures. 

\begin{figure} [hbtp]
  \centering
  \includegraphics[scale=0.76]{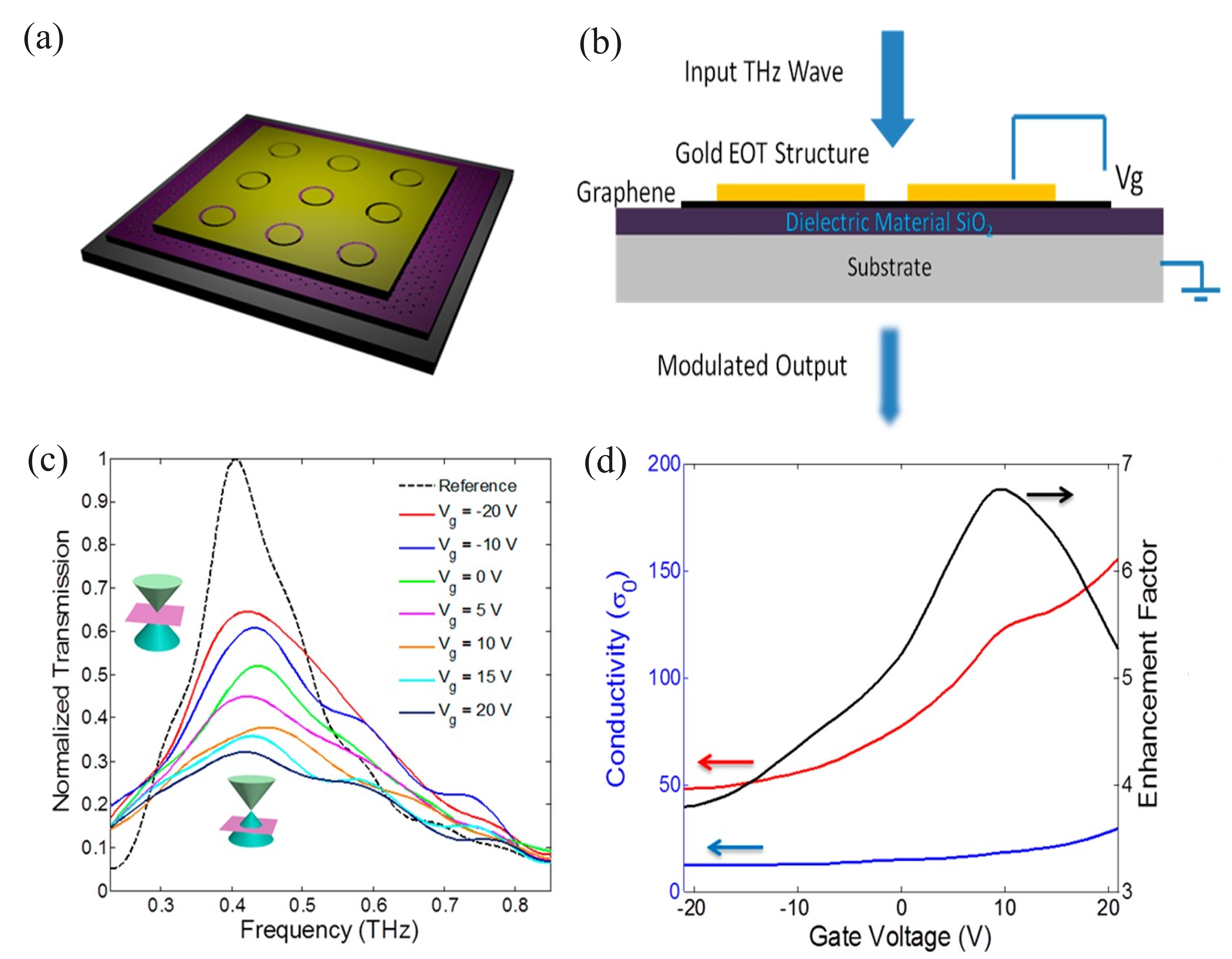}
  \caption{EOT graphene-based THz modulator. (a)~System schematic. (b)~Cross-sectional view of the modulator. The EOT structure is positioned above graphene, which is placed on a SiO$_2$/Si substrate. (c)~Transmission spectra for the modulator with gate voltages ranging from $-$20 to $+$20\,V. The dashed curve represents the reference transmission measured for the EOT structure on SiO$_2$ without graphene. (d)~Real part of the conductivity of the monolayer graphene with (red line) and without (blue line) the EOT structure at the resonance frequency along with the enhancement factor (black line) against the gate voltage. Reproduced with permission from \cite{gao2014high}. Copyright 2014, American Chemical Society. }
  \label{fig:9}
\end{figure}

In the study described in~\cite{gao2014high}, the structure of the graphene-assembled modulator comprises an EOT resonator consisting of a patterned lattice of ring apertures that operate in the THz range. The resonator is sandwiched on a $p$-doped graphene layer, which is placed above the dielectric SiO$_2$ substrate~\cite{suk2011transfer}; see \textbf{Figures~\ref{fig:9}}(a) and (b). To modulate the carrier density, a gate voltage is applied between the lower silicon substrate and the upper EOT structure. The ring aperture is subjected to a normally incident THz wave with linear polarization that induced a bright dipole mode. The radial electric-field element of the mode experiences a 2$\pi$ variation in its phase along the ring contour, leading to a resonant EOT effect. This resonance occurs at frequency $f_0 = c/(2\pi rn_\mathrm{eff})$, where $r$ is the ring radius and $n_\mathrm{eff}$ is the effective index of the mode. The monolayer graphene absorption increases due to near-field enhancement resulting from EOT. The absorption enhancement factor is expressed as the ratio of the absorption in the presence of the EOT structure to the absorption in its absence. The enhancement is shown to be proportional to the monolayer graphene near-field intensity~\cite{gao2014high}. 

Transmission spectra for the THz modulator based on EOT graphene were normalized using the reference transmission spectrum corresponding to the EOT structure minus the graphene monolayer. The transmission spectra were obtained through THz-TDS while varying $V_\mathrm{g}$ within the range of $-$20 to $+$20\,V. They showed insensitivity to polarization, in accordance with the structural symmetry. In Figure~\ref{fig:9}(c), the central frequency showed minimal change across different gate voltages, whereas the transmission peak changed in response to the gate voltage. The variation in transmission is a consequence of the change in carrier concentration within graphene, which occurs due to the shift in $E_\mathrm{F}$ induced by $V_\mathrm{g}$~\cite{ren2012infrared}.

In Figure~\ref{fig:9}(d), the real part of the conductivity (in units of $\sigma_0$\, representing the universal interband optical conductivity~\cite{klimchitskaya2016conductivity}) and the enhancement factor are displayed as a function of $V_\mathrm{g}$. At the resonance frequency, the EOT structure exhibits a conductivity (red line) approximately seven times greater than that observed in the absence of the structure (blue line). The primary contributor to this improvement is the robust near-field enhancement around the ring apertures, amplifying the interaction between the THz field and the monolayer graphene~\cite{gao2014high}.
The breakdown of the SiO$_2$ layer limits the gate voltage to a maximum reverse bias voltage of $-$20 V, which is when the THz absorption of graphene reaches a nonzero value, as indicated by the red line in Figure~\ref{fig:9}(d). This suggests that even when $E_\mathrm{F}$ gets closer to the Dirac point, graphene will continue to exhibit a measurable THz conductivity, which is ascribed to the nonuniformity in $E_\mathrm{F}$ across extensive areas of graphene~\cite{buron2012graphene}. Therefore, residual THz loss is induced by the presence of residual carriers in the film, as the Dirac point cannot be attained by the entire area under the same applied voltage~\cite{ren2012infrared, gao2014high}.

Additionally, the study in~\cite{sensale-rodriguez_broadband_2012} highlights a remarkable improvement in THz wave modulation, demonstrating a modulation capability exceeding 2.5 times that of previous broadband intensity modulators. This was achieved through the development of a prototype graphene THz modulator. The modulator consists of monolayer graphene positioned on a SiO$_2$/$p$-Si substrate, featuring two top contacts to measure graphene conductivity and a bottom ring gate contact to adjust the graphene Fermi level. The modulation of THz waves, resulting from alterations in absorption within monolayer graphene, leads to impressive tuning from 5\% to 20\%. Further, recent advancements in gate-tuned graphene meta-devices have expanded the landscape of THz wave manipulation. Dynamic control of THz wavefronts has been achieved using gated ion gel and graphene metadevices~\cite{Li2022gate}. Graphene metasurfaces have also been used to modulate the amplitude of refracted THz waves~\cite{Kim2017Amplitude}, reflected THz waves~\cite{squires_electrically_2022}, and tunable phase~\cite{Miao2015Widely}. Low-bias THz amplitude modulation has also been accomplished using a graphene and split-ring resonator (SRR) metasurface~\cite{valmorra_low-bias_2013,Delinnocenti2014Low}, an H-shaped gold metamaterial~\cite{jung_electrically_2018}, a metallic spiral microstructure and an all-dielectric metamaterial~\cite{yao_frequency-dependent_2021}. A graphene/quartz modulator demonstrated a further ability to tune the Brewster angle~\cite{Chen2018Graphene}. In addition to electrostatic gating, several all-optical controlled graphene metasurface THz amplitude modulators have been proposed~\cite{fan_photoexcited_2018,tasolamprou2019experimental,choi_augmented_2021}. The depth rate of amplitude modulation of graphene metamaterial devices has significantly improved in recent years~\cite{wang_dynamic_2022}. These recent studies highlight the ongoing progress in graphene-based THz modulation technologies.


\subsection{Phase Modulation for Communication}

In the preceding subsections on THz-wave modulation by graphene, we centered our discussion on the principles of gating and its application in modulating the amplitudes of THz waves. We explored strategies to enhance absorption and hence the modulation depth through advanced measurement techniques and metasurface-assisted architectures. Building upon these foundational discussions, this section aims to delve deeper into the synergistic integration of these diverse techniques. Specifically, we investigate how their combination could facilitate the realization of phase, waveform, and polarization modulations, paving the way for developing advanced THz communication devices.

Phase modulation of EM waves is a key area of interest in photonics, primarily because manipulating the local EM phase enables advancements in applications such as wavefront control~\cite{yu2011light}, holography~\cite{yifat_highly_2014,zheng_metasurface_2015,DENG201716,jiang2019metasurface}, and imaging~\cite{zhang_reconfigurable_2021}.
Achieving phase modulations that span the full $360^\circ$ range can be accomplished using metasurfaces~\cite{wang2021full}. Nevertheless, metasurface-based phase modulators are typically passive and lack the ability to be externally tuned~\cite{gong2022active}. On the other hand, using gate-controlled graphene has shown potential in adjusting the resonance characteristics in photonic systems, though achieving only limited phase-tuning ranges~\cite{ju2011graphene,sensale2012extraordinary,yao2014electrically}.

\begin{figure}[hbtp]
  \centering
  \includegraphics[scale=0.76]{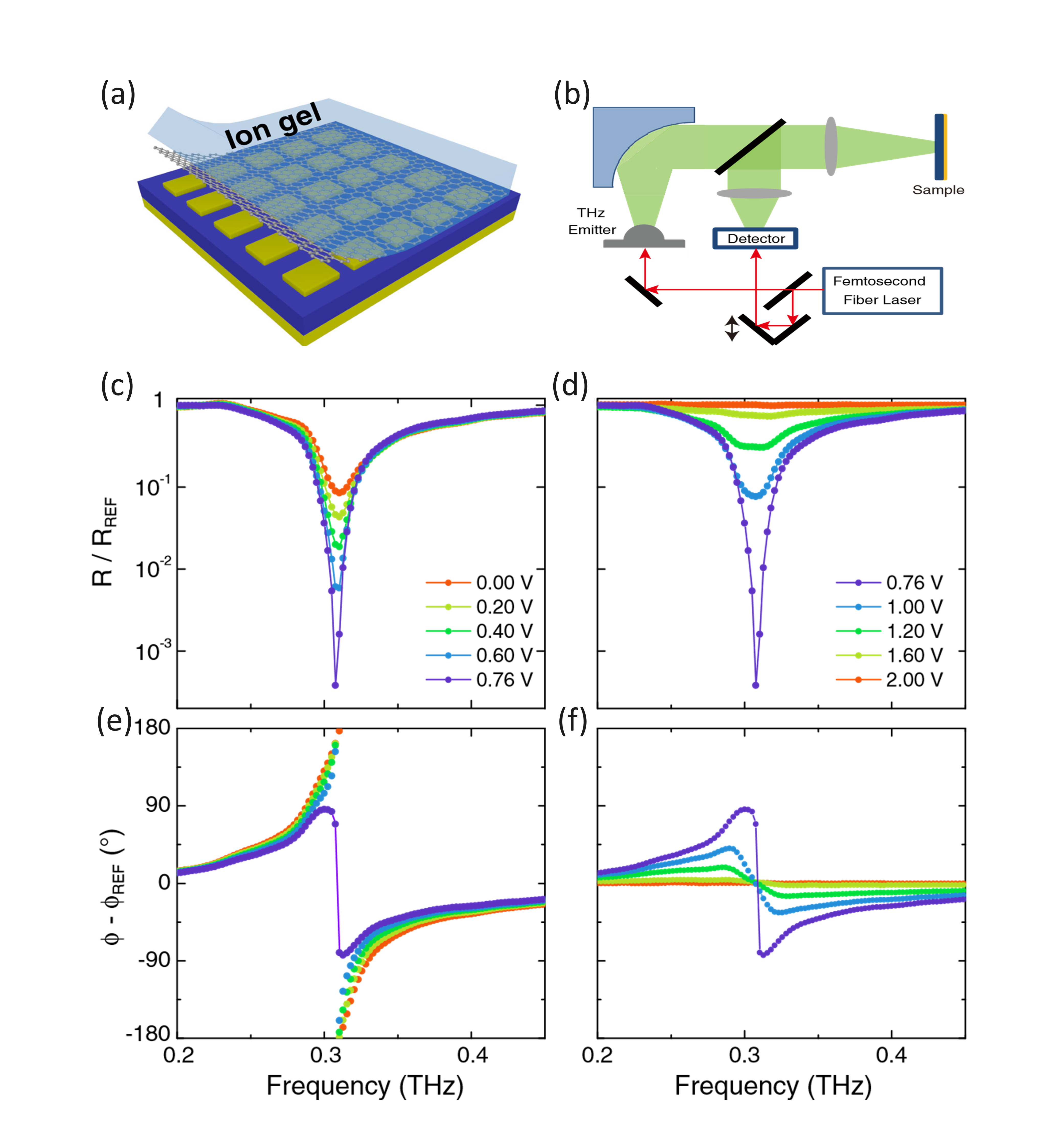}
  \caption{Phase modulation with a graphene metasystem. (a)~Representation of the graphene-based  metasurface structure. (b)~Illustration of the THz time-domain spectroscopy measurement setup. (c-d)~The reflectances for $\Delta V_\mathrm{g}\le \Delta V_\mathrm{C}$ ($\Delta V_\mathrm{g}\ge\Delta V_\mathrm{C}$). The spectrum obtained at $\Delta V_\mathrm{g}$= 2.02\,V is used as a reference. (e-f)~The phase modulations for $\Delta V_\mathrm{g}\le \Delta V_\mathrm{C}$ ($\Delta V_\mathrm{g}\ge\Delta V_\mathrm{C}$). Reproduced with permission from \cite{Miao2015Widely}. Copyright 2015, American Physical Society.}
  \label{fig:phase1}
\end{figure}

A recent study~\cite{Miao2015Widely} has introduced another approach that extends the phase-modulation range by employing an ultrathin metasystem coupled with ionic-gated monolayer graphene. When a gate voltage is applied to graphene, its optical conductivity changes, as described in Section~\ref{2.1}. This modification alters the metasurface/graphene system from an underdamped to an overdamped resonator, substantially modulating the phase of the reflected wave.
The system comprises a one-port resonator structure, which is capable of shifting the phase of the reflected wave through a $180^\circ$ transition by precisely controlling the losses via the graphene layer. As shown in \textbf{Figure \ref{fig:phase1}}(a), the metasurface consists of a SiO$_2$/Si substrate, an aluminum film, an SU8 spacer, aluminum mesas, graphene, and gel-like ion liquid. A gate voltage $V_\mathrm{g}$ is applied between the graphene layer and an upper gate to modulate the graphene's resistance and, consequently, the resonator's loss.

\begin{figure}[hbtp]
  \centering
  \includegraphics[scale=0.66]{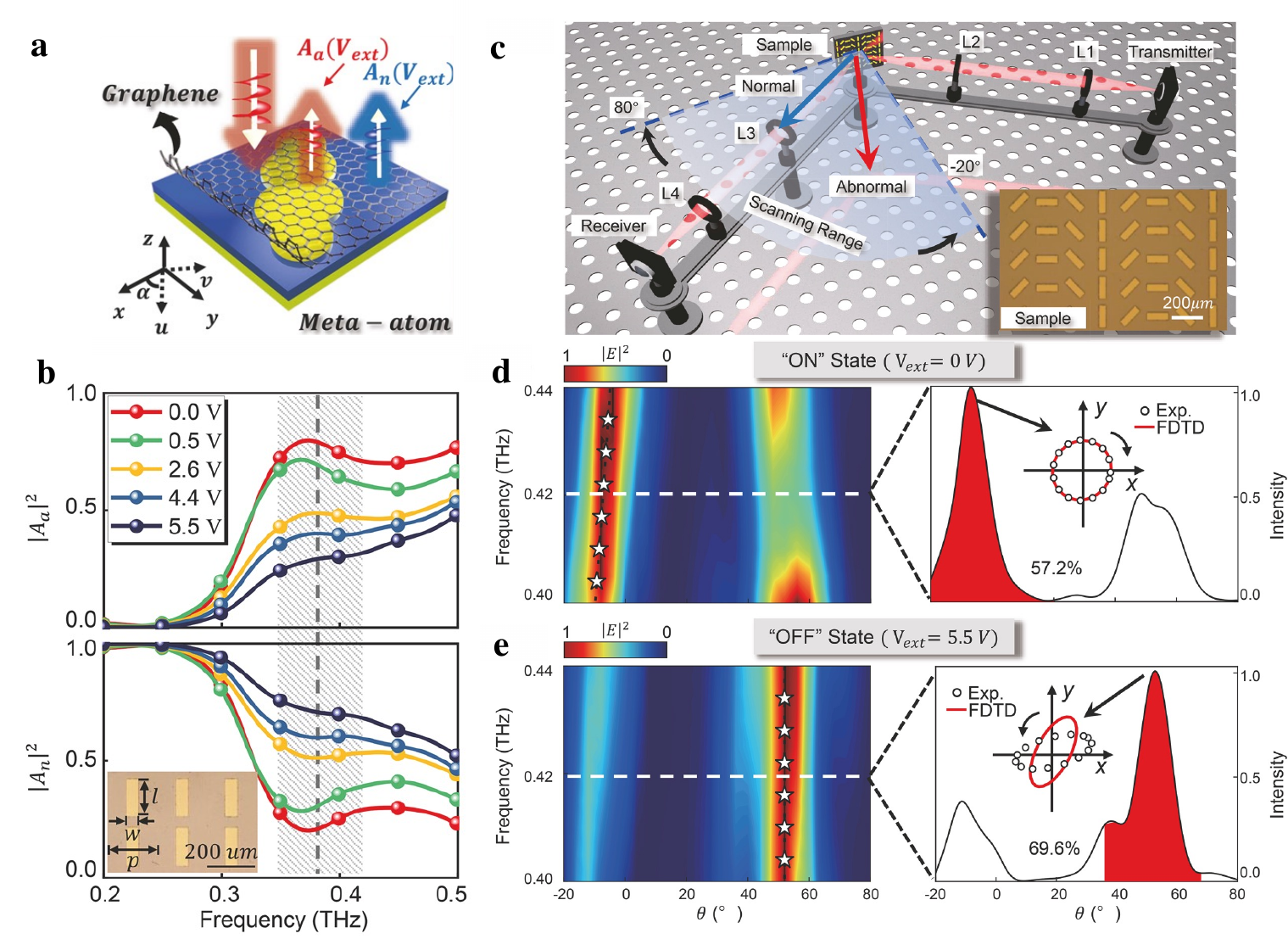}
  \caption{Gated-graphene meta-deflector for dynamic tunability of EM wavefronts. (a)~Schematic of the meta-atom unit cell system consisting of a sub-wavelength patterned metal structure with graphene on top. $\mathbf{u}$ and $\mathbf{v}$ are the principal axes of the metasurface. Scattered beams consist of abnormal and normal modes. (b)~Frequency-dependent $|A_\mathrm{a}|^{2}$ abnormal-mode (top) and normal-mode $|A_\mathrm{n}|^{2}$ (bottom) intensities at various $V_\mathrm{ext}$. Inset: optical image  of the metasurface. (c)~Angle-resolved THz-TDS setup showing two deflected beams. Inset: Optical image of the designed metasurface. (d-e)~Left: Scattering electric field intensity color map of the reflected ON (OFF) state at $V_\mathrm{ext}$ = 0\,V $(V_\mathrm{ext}$ = 5.5\,V) as a function of detected deflection angles and measured THz frequency, right: cross section of (left) at 0.42\,THz with deflection efficiency annotation and polarization state representation, respectively. Reproduced with permission from \cite{Li2022gate}. Copyright 2022, American Physical Society.}
  \label{fig:phase2}
\end{figure}

The reflection amplitude and phase are analyzed using a THz-TDS system; see Figure \ref{fig:phase1}(b).
The reflectances are depicted in Figures \ref{fig:phase1}(c) and (d), and the corresponding phase spectra in Figure \ref{fig:phase1}(e) and (f), at various gate voltages. 
The doping level of graphene is determined by the gate voltage relative to the Dirac point $\Delta V_\mathrm{g} = V_\mathrm{g} - V_\mathrm{D}$, with $V_\mathrm{D}$ being the charge-neutrality point.
At the maximum doping level ($\Delta V_\mathrm{g}$ = 2.02\,V), the spectrum exhibits no discernible features and is used as a reference.
At the resonance frequency $f$ = 0.31\,THz, a dip in reflectance is noted, accompanied by a consistent $360^\circ$ phase shift across the frequency, typical of a magnetic resonator's intrinsic behavior~\cite{hao2007manipulating}.
When the doping level increases on the electron side, $\Delta V_\mathrm{g} > 0$, the reflectance at resonance steadily decreases. Despite a narrowed bandwidth [Figure \ref{fig:phase1}(e)], the resonator still demonstrates its magnetic response, exhibiting a $360^\circ$ phase variation.
Nonetheless, increasing the gate voltage past a critical value ($\Delta V_\mathrm{C}$ = 0.76\,V) reverses this trend, leading to increased reflectance at resonance; see Figure \ref{fig:phase1}(d). The phase spectrum ceases to display a magnetic-resonance trait and instead demonstrates a diminishing phase variation around $0^\circ$. These results not only illustrate the $180^\circ$ phase modulation but also mark a critical transition at $\Delta V_\mathrm{g}$= $\Delta V_\mathrm{C}$ in the metasystem's response, a behavior also observed when doping on the hole side ($\Delta V_\mathrm{g} < 0$). The authors also report a larger phase modulation by employing two graphene metasurfaces gated independently. In a similar independent study~\cite{Miao2015Widely}, an enhanced modulation range of up to $243^\circ$ has been reported. Such tunable graphene metasurfaces show potential across a multitude of applications, including the independent gating of phase bits to manipulate THz wavefronts.

As an extension to the preceding study, further advancement in manipulation of polarization and deflection of reflected beams has been reported~\cite{Li2022gate}. 
The authors devised a similar device to the previous one, comprising the following sandwich: SiO$_2$/Si substrate, metal, insulator, monolayer graphene, metasurface, and ionic gel, integrating subwavelength metasurface features and graphene synthesized via CVD [\textbf{Figure \ref{fig:phase2}}(a)]. Voltage $V_\mathrm{g}$ is applied between the graphene layer and the top ionic gate to modulate absorption. 
Optical spectroscopy involved illuminating the sample with linearly polarized light at normal incidence, with configurations aligned along the two principal axes of the metasurface, followed by time-domain signal measurements.
The external voltage $V_\mathrm{ext}$ is controlled as $V_\mathrm{D}-V_\mathrm{g}$. By varying $V_\mathrm{ext}$, the authors successfully manipulated the intensities of abnormal-mode $|A_\mathrm{a}|^{2}$ and the normal-mode $|A_\mathrm{n}|^{2}$ (at 0.38\,THz). Upon increasing the gating voltage from 0 to 5.5\,V, a significant decrease in abnormal-mode intensity and a notable increase in normal-mode intensity were observed [Figure \ref{fig:phase2}(b)]. Notably, a crossing of both components occurs at $V_\mathrm{ext}$ = 2.6\,V.

The graphene layer acts as a tunable lossy medium under gating, and the absorption loss is highly influenced by the geometric parameters of the metasurface, resulting in distinct reflection coefficients along the two principal axes of the patterned metasurface. As $V_\mathrm{ext}$ increases, the device's scattered far field reflection transitions from being abnormal mode-dominated to normal mode-dominated. Furthermore, angle-resolved TDS measurements were conducted by illuminating the sample with left-handed circularly polarized incident light at a $50\degree$ angle [Figure \ref{fig:phase2}(c)], where, according to Snell's law, deflection of normal and abnormal modes occurs at $50\degree$ and $-7.3\degree$, respectively. The device was characterized under $V_\mathrm{ext}$ = 0\,V and 5.5\,V. At null gating voltage, two scattering regions appear: at 0.42\,THz, the reflected beam is abnormal mode-dominated, while at $V$ = 5.5\,V, it is normal mode-dominated. As depicted by Figure \ref{fig:phase2}(d) and (e), by rotating the linearly polarized transmitter and receiver, the authors observed left-handed circular polarization (denoted by the ON state) for $V = 0$ (efficiency: $57.2$) and right-handed circular polarization (denoted by the OFF state) for $V$ = 5.5\,V (efficiency: $69.9$).

Several other graphene metamaterial devices with different substrate or structure design can control both the amplitude and phase of the transmitted THz wave, such as a hybrid metamaterial system consisting of SRRs capacitively coupled to the graphene layers~\cite{balci_electrically_2018} or a graphene-assisted meta-lens~\cite{liu_graphene-enabled_2018}, or SRRs made of graphene on a magnesium fluoride substrate~\cite{chen_novel_2019}. The amplitude and phase of the reflected THz wave can also be tuned by different graphene metasurface structures, as shown in~\cite{kakenov_graphene_2018,zhang_novel_2018}. A graphene metamaterial on a GaAs substrate can enable all-optical control of THz amplitude transmission, phase and group delay, and polarization conversion, utilizing a dual metamaterial induced transparency (dual-MIT) phenomenon~\cite{deng_multifunctional_2023}. There is also an experimental demonstration that integrates graphene metamaterial with perovskite into a metadevice that modulates THz amplitude and phase via a bias voltage or optical pumping~\cite{yang_dual-stimulus_2022}.
Further, several graphene-based metasurface THz polarization converters (quarter-wave plates and/or half-wave plates) have been demonstrated~\cite{guo_broadband_2016, guan_bi-functional_2019, zhang_bi-functional_2020, fu_tunable_2021, zhang_broadband_2021,park_electrically_2023,deng_multifunctional_2023}. These converters are broadband and easily tunable by electrostatic gating.
These reported results are promising for realizing next-generation communication systems such as THz radars and vectorial beam coded protocols.
\section{Conclusions and Prospects}
In this review, we summarized some distinctive characteristics of graphene and explored their benefits for applications in THz technology for sensing and communication. Studying the adsorption and desorption of O$_2$ molecules on graphene under different conditions has helped explain the variations in emitted THz waves. This interaction between external molecules and graphene has enabled the development of flexible graphene devices capable of detecting pesticides on biointerfaces without needing metasurfaces. In a microfluidic cell, DNA detection has been realized at high sensitivity and specificity, enabling \textit{in~situ} THz sensing of liquids. Furthermore, the modulation of THz transmission has been successfully achieved by tuning the Fermi level, leading to manipulating the carrier density with an external gate voltage. Parallel-plate waveguides have proven effective in modulating transmitted THz waves, exhibiting better sensitivity than traditional THz spectroscopy. The phenomenon of total internal reflection in a unique waveguide structure has revealed that the presence of graphene results in enhanced absorption. The use of metallic ring apertures has demonstrated the enhancement of THz absorption in graphene through the extraordinary optical transmission phenomenon. By combining gating with metasurface fabrication, graphene has shown substantial THz phase modulation, with the ability to reflect incident THz waves toward determined directions with controlled polarization upon gating or optical pumping. Due to the unique combination of exceptional physical properties of graphene and its versatility, many technological limitations present in conventional devices operating in the THz frequency range can be overcome, making it suitable for a wide range of applications from sensing to wireless communication. Hence, upon all the reported findings, there is a considerable opportunity for further game-changing research studies in this field.

\section*{Acknowledgements}
We acknowledge support from the Robert A.\ Welch Foundation through Grant No.\ C-1509, the Air Force Office of Scientific Research through Grant No.\ FA9550-22-1-0382, and the Chan Zuckerberg Initiative through Grant No.\ WU-21-357.

\medskip

\bibliography{BIBLIOGRAPHY}

\end{document}